\documentclass[12pt,superscriptaddress,preprint]{revtex4-1}
\usepackage{graphicx}
\usepackage{epstopdf}
\usepackage{amsfonts}
\usepackage{amssymb}
\usepackage{amsthm}
\usepackage{amsmath}
\usepackage{multirow}

\newcommand{\newc}{\newcommand}
\newc{\ra}{\rightarrow}
\newc{\lra}{\leftrightarrow}
\newc{\be}{\begin{equation}}
\newc{\ee}{\end{equation}}
\newc{\bs}{\begin{split}}
\newc{\es}{\end{split}}
\newc{\ba}{\begin{eqnarray}}
\newc{\ea}{\end{eqnarray}}
\newc{\ov}{\overline}
\newc{\pa}{\partial}
\newc{\D}{\Delta}

\newc{\nn}{\nonumber}

\begin{document}

\vspace*{0.75in}

\title{Magnetic Monopoles and Free Fractionally Charged States at Accelerators and in Cosmic Rays}

\author{Thomas W. Kephart}
\email{tom.kephart@gmail.com}
\affiliation{Department of Physics and Astronomy, Vanderbilt
University, Nashville, TN 37235 USA}

\author{George K. Leontaris }
\email{leonta@uoi.gr}
\affiliation{Physics Department, Theory Division, Ioannina University,
GR-45110 Ioannina, Greece}

\author{Qaisar Shafi}
\email{shafi@bartol.udel.edu}
\affiliation{Bartol Research Institute,
University of Delaware, Newark, DE 19716 USA}

\date{\today}

\begin{abstract}
Unified theories of strong, weak and electromagnetic interactions which have electric charge quantization predict the existence of topologically stable magnetic monopoles.
Intermediate scale monopoles are comparable with detection energies of cosmic ray monopoles at IceCube and other cosmic ray experiments. Magnetic monopoles in some models can be significantly lighter and carry two, three or possibly even higher quanta of the Dirac magnetic charge. They could  be light enough for their effects to be detected at the LHC either directly or indirectly. An example based on a D-brane inspired $SU(3)_C\times SU(3)_L\times SU(3)_R$ (trinification) model with the monopole
carrying three quanta  of Dirac magnetic charge is presented. These theories  also predict the existence of color singlet states with fractional electric charge which may be accessible at the LHC.

\end{abstract}


\maketitle

\section{Introduction}
The fact that electric charge is quantized lead Dirac~\cite{Dirac} in 1931 to predict the existence of magnetic monopoles.
Classically, a stationary system consisting of a magnetic monopole and an electron has a non-vanishing Poynting vector and angular momentum. Quantum mechanically, angular momentum must be quantized in units of $\hbar$ and this implies the Dirac quantization condition (in units where $\hbar=c=1$)
\[gq=2n\pi\,,\]
where $q$ is the electric charge, $g$ is the magnetic   charge and $n$ is an integer.
Dirac's argument is still compelling today but magnetic monopoles have eluded  us after over eighty years of searching. The discovery of magnetic monopoles would have wide reaching implication for physics beyond the standard model. As a new energy regime has been opening up at the Large Hadron Collider (LHC), it is important to be clear on what we expect could be found as we extend the search for magnetic monopoles into this region. While some results are expected to be model dependent, others will be universal. We discuss a class of models that could have magnetic monopoles light enough to have implications for the LHC,
as well as heavier monopoles that may be  observed in cosmic ray experiments.

\section{Generalities for Product Group Models}

Two familiar examples of product gauge groups with bifundamental fermions are the Pati-Salam (PS) model \cite{PS} and the trinification model \cite{Tri}.
In the PS model the gauge group is $SU(4)\times SU(2)\times SU(2)$  and the fermions live in  three $[(4,2,1)+(\bar4,1,2)]$ families which reduces to three standard model (SM) families plus three right handed neutrinos. This model can be embedded directly into $SO(10)$ with no additional fermions. In trinification models the gauge group is $SU(3)\times SU(3)\times SU(3)$ and the fermions occupy  three $[(3,\bar3,1)+(1,3,\bar3)+(\bar3,1,3)]$ families which reduces to three standard model (SM) families plus the additional content of three $E_6$ families, including additional $b$-type quarks. Let us begin by discussing the magnetic monopoles of these two models and their generalizations.

\subsection{Pati-Salam model}

It has been understood for some time that the spontaneous breaking of the Pati-Salam gauge symmetry $H = SU(4)_C \times SU(2)_L \times SU(2)_R$ (422 model) yields topologically stable monopoles that carry two quanta of Dirac magnetic charge \cite{Lazarides:1980cc,Shafi:2007xp,Lazarides:1982jq}. By not insisting that $H$ be embedded within an $SO(10)$ model, this implies that in this model there should exist $SU(3)_c$ color singlet states that carry fractional $(\pm \frac{e}{2})$ electric charge. Adding fundamental fermions irreducible representations  to the 422 model is an obvious extension. For instance, since $SU(2)$ is anomaly free, and as the 422 gauge group has no $U(1)$ factors, we could simply
introduce $(1,2,1)$ and $(1,1,2)$ states in the fundamental representations of $H$ which provide the required $SU(3)_c$ singlet states that carry fractional charge. (Recall that the known fermions belong in the bi-fundamental representations of $H$.) Moreover, we also should include the conjugate pair $(4,1,1)$ and $(\bar4,1,1)$ in the fundamental representations of $H$, which transform as triplets and anti-triplets under $SU(3)_c$ and carry fractional charge $\pm \frac{e}{6}$.  These latter states could bind together to create, for instance, a new class of baryons that carry electric charge $\pm \frac{e}{2}$. They also could combine with the SM quarks to generate fractionally charged hadrons. This leads to color singlet magnetic monopoles carrying integer multiples of the Dirac charge \cite{Lazarides:1982jq}, in this case $g=\pm \frac{2e}{2\alpha}$.   In principle, the scale of the new fermions can be arranged to be light, perhaps even LHC accessible.
The monopole mass depends, of course, on the 422 breaking scale.  Intermediate mass monopoles may survive inflation as we will discuss.

\subsection{Trinification model}
As the gauge group for the trinification model (333 model) is $H = SU(3)_C \times SU(3)_L \times SU(3)_R$ and
since all the $SU(3)$s are potentially anomalous, the simplest generalization is to add fundamental fermion  representations in conjugate pairs, e.g., $(3,1,1)+ (\bar3,1,1)$. There is also the possibility of adding combinations of fundamentals and bifundamentals that cancel the anomaly. For instance, we could add $[3(3,1,1)+ 3(1,\bar3,1)]+(\bar3,3,1)]$ and the theory would remain free of chiral anomalies. The additional fundamental fermions lead to leptons with electric charges $\pm\frac{2e}{3}$ and hence charge $\pm 3e$ magnetic monopoles. 

\subsection{Lowering the GUT scale}
If all the gauge coupling constants of a product group start off equal at the GUT scale \cite{GQW,Georgi:1974sy}, then we expect the GUT scale to be rather high, $M_{U}\sim 10^{16}$ GeV. However, there are cases where equality at the GUT scale is not required. For instance, in orbifolded AdS$_5$/$S^5$ with abelian orbifolding group $Z_n$ and gauge group $SU(3)^n$ one finds that the gauge group coupling constants can be related by rational fractions. For trinification models the ratios are determined by how the three $SU(3)$s are diagonally embedded into the initial $SU(3)^n$ group. (See \cite{Lawrence:1998ja} and the detailed discussion in \cite{Ho:2011qi}.) This then allows the GUT scale to be considerably lower since less RG running is required for unification.

Another way to lower or alter the GUT scale is by adding extra dimensions to allow power law running of couplings \cite{extraD}. Yet another is to add scalar thresholds~\cite{Dvali:1994wj,Berezhiani:2001ub,Barr:2014nza} or vector-like fermion thresholds. All these methods can be arranged to avoid proton decay at a too rapid rate. For the remainder of this work we will assume one of these mechanisms is operating to avoid proton decay and lower the GUT scale. This will allow the GUT symmetry to break and $U(1)$ factors to appear at a low scale, which in turn delivers light magnetic monopoles with charges depending on the gauge group and fermionic content of the model.

\subsection{  The 433 model}
Now let us consider  extensions of the Pati-Salam and trinification models that naturally contain both fundamental and bifundamental  representations. The simplest case is based on the gauge group $SU(4)\times  SU(3)\times SU(3)$ (433 model), where both the 422 and  trinification models can be embedded \cite{Kephart:2001ix}. However, these are not the only possibilities. In all there are 18 inequivalent embeddings \cite{Kephart:2006zd} of the standard model gauge group in $SU(4)\times  SU(3)\times SU(3)$. If we insist on bifundamental fermions at the 433 level, then  the 433 model is only anomaly free when families come in a multiple of three.
 At the 422 and trinification level, the 433 model naturally delivers both fundamental and bifundamental fermions. Hence fractional electric charge color singlets and multiply charged magnetic monopoles are natural in the 433 model.

 Our main objective here is to find allowed masses and charges of magnetic monopoles and then suggest signatures for experimental searches. The 433 model is a good candidate for a model that can have detectable multicharged magnetic monopoles. (Here we focus only on models similar to the extended versions of the 422 and 333  models derivable from the 433 model and will save the exploration of the full set of non equivalently embedded SMs for further work.) The magnetic monopole spectrum for the extended versions of the 422 and 333 models   under various model assumptions  \cite{Cho:1996qd,Cho:2012bq,Vento:2013jua,Cho:2013vla}  then suggest where experimental searches may have the best chance of success.

 \section{Trinification from intersecting D-brane  scenario with Observable Monopoles}

 In this section we explore a string motivated trinification model with monopoles that can be light enough to be observed, in future colliders 
 as well as ongoing cosmic ray searches.
More specifically, we will present an  interesting supersymmetric  version which is realised in the framework of intersecting D-branes.
 We will  describe here the  basic steps  for such a viable D-brane construction.  The trinification group is generated by
three stacks of D-branes, each stack containing three parallel almost coincident branes. Each stack gives rise to a
$U(3)$ gauge group which results in  the gauge symmetry~\cite{Anastasopoulos:2006da,Leontaris:2008mm}
\be
U(3)_C\times U(3)_L\times U(3)_R\,.
\ee
In this notation,  the first $U(3)$ contains  the  $SU(3)$ color group of the SM gauge symmetry,
the second $U(3)$ includes the weak $SU(2)_L$, and the  third $U(3)$ contains  the $SU(2)_R$ gauge group. From the group relation
$U(3)\simeq SU(3)\times U(1)/Z_3$,  in addition to the standard
$SU(3)^3$  trinification gauge symmetry, the D-brane analogue is augmented by three extra $U(1)$'s.
The final local symmetry can be written
\ba
SU(3)_c\times SU(3)_L\times SU(3)_R\times U(1)_C\times
U(1)_L\times U(1)_R\,.\label{333111}
\ea
The abelian $U(1)_{C,L,R}$ factors have mixed anomalies with the non-abelian $SU(3)^3$ gauge part, but there is
an anomaly free combination,
\ba
U(1)_{{\cal Z}'}&=& U(1)_C+U(1)_L+U(1)_R\label{aas}\,\cdot
\ea
The anomalies associated with the two remaining combinations are cancelled by a generalized Green-Schwarz mechanism
and the corresponding  bosons receive masses from four-dimensional couplings 
involving  the Ramond-Ramond scalars coming from the
twisted closed string spectrum~\cite{Ibanez:1998qp}. Furthermore,  there is a remaining global  symmetry
 associated with $U(1)_C$ of the color gauge group factor $U(3)_C\simeq SU(3)_C\times U(1)_C/Z_3$, which 
can be  identified with baryon number that is conserved at the perturbative level.

\subsection{Spectrum}

Next we briefly  present the salient features of the spectrum. In intersecting D-branes the fermion and Higgs fields
are generated by open strings with ends attached either on the same brane stack, or on two different brane stacks. 
In the most general picture (as in the presence of orientifolds),
there are also strings with one end attached on mirror brane stacks giving rise to additional states. More precisely,
open strings with ends on two different brane stacks give rise to bifundamentals, while strings with both ends on the
same (or with one end on a mirror) stack   introduce, among others, adjoint, antisymmetric  and singlet representations.
For the trinification model in particular, the bifundamentals are of the well-known form
$(3,\bar 3,1)$, $(1,3,\bar 3)$ and $(\bar 3, 1,3)$.

\noindent
Additional representations corresponding to open strings with ends on the same (or mirror) stacks
may appear in the massless spectrum. These transform only under one gauge factor and they are formed
according to  $3\times 3=\bar 3+6$ and $3\times\bar 3= 1+8$.
 Note that for  the $SU(3)^3$ symmetry, in particular,  these can generate states in $(3,1,1)$, $(1,3,1)$, and  $(1,1,3)$.
All of these states are `charged' under the $U(1)$ factors.

The standard  matter representations of the 333 model arise from strings with ends on different brane stacks
 and have the quantum numbers
\ba
{\cal Q^{\hphantom{c}}}&=&(3,\bar 3,1)_{(+1,-1,\hphantom{+}0)}\label{QL}\\
{\cal Q}^c&=&(\bar 3,1,3)_{(-1,\hphantom{+}0,+1)}\label{QR}\\
{\cal L}^{\hphantom{c}}&=&(1,3,\bar
3)_{(\hphantom{+}0,+1,-1)}\,,\label{LH}
\ea
The three lower indices refer to the three abelian factors $U(1)_{C,L,R}$  discussed above.
The Higgs content may be accommodated   in the  bifundamentals
\ba
{\cal H}_a &=&(1,3,\bar
3)_{(\hphantom{+}0,+1,-1)},\;a=1,2\,.
\ea
There are also  representations generated with both ends on
the same brane  stack, such as
\ba
{\cal H}_{\cal L}&=&(1,3,1)_{(0,-2,0)}\label{HL}
\\
{\cal H}_{\cal R}&=&(1,1,3)_{(0,0,-2)}\label{HR}
\\
{\cal H}_{\cal C}&=&(3,1,1)_{(-2,0,0)}\label{HCC}\,,
\ea
and their complex conjugates (c.c.).
\begin{figure}
\includegraphics[width=70mm]{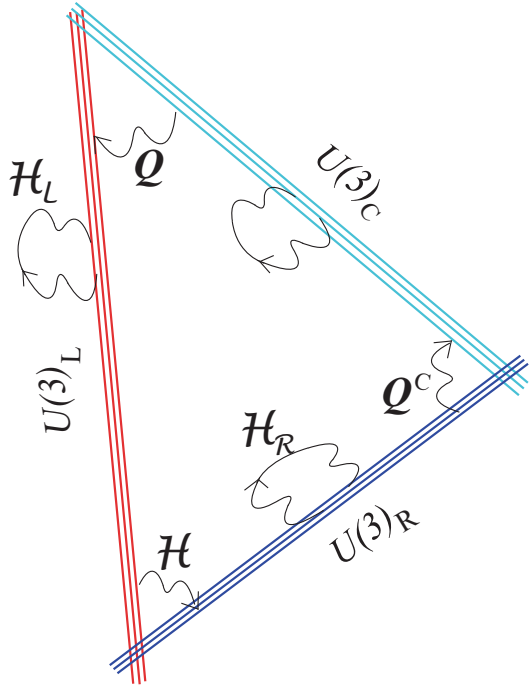}
\caption{Intersecting D-brane stacks for the trinification model.
Also shown are strings attached to D-branes whose excitations give rise to the
representations explained in the text.}
\end{figure}

Under $SU(3)_C\times SU(2)_L\times
U(1)_Y\times U(1)_{\Omega}$ (where  $U(1)_{\Omega}$ is left over from
the $SU(3)_R$ breaking)  the MSSM states have the following assignments
\ba
{\cal Q}^{\hphantom{c}}&=&
q\left(3,2;\frac{1}{6},0\right)+g\left(3,1;-\frac{1}{3},0\right)\nonumber\\
{\cal Q}^c&=&
d^c\left(\bar3,1;\frac{1}{3},1\right)+u^c\left(\bar3,1;-\frac{2}{3},0\right)
+g^c\left(\bar3,1;\frac{1}{3},-1\right)\label{fermions}\\
{\cal L}^{\hphantom{c}} &=&
\ell^+\left(1,2;-\frac{1}{2},1\right)+\ell^-\left(1,2;-\frac{1}{2},-1\right)
+{\ell^c}\left(1,2;+\frac{1}{2},0\right)\nonumber\\&~&+\;
\nu^{c+}(1,1;0,1)+\nu^{c-}(1,1;0,-1)+e^c(1,1;1,0)~,
 \nonumber
\ea
and similarly for the Higgs scalars ${\cal H}_a+c.c.$

The `standard'  hypercharge assignment corresponds to  a linear combination of the
$U(1)$ generators $X_L'$ and $X_R'$ of $SU(3)_L$ and  $SU(3)_R$ respectively
\ba
Y=-\frac{1}{6} X_{L'}+\frac 13 X_{R'}~\cdot\label{SY}
\ea
Under the above hypercharge embedding, all MSSM particles obtained from the decompositions
in~(\ref{fermions}) acquire their SM charges. 
However, we have observed that additional superfields are also available  from strings with both ends attached on the 
same brane stack and  under the  hypercharge assignment~(\ref{SY}), they are fractionally charged.
The electric  charges of the $SU(2)_L$ triplet components ${\cal H}_L=(1,3,1)+c.c.$,  in particular, are found to be
fractional $\pm \frac e3,\,\pm \frac{2e}3$.

We have seen already that in the present D-brane construction,  the three additional abelian factors define
the anomaly free linear combination~(\ref{aas})
which can be used  to redefine the hypercharge according to
\ba
Y'=Y+\frac 16{\cal Z}'=-\frac{1}{6} X_{L'}+\frac 13 X_{R'}+\frac 16{\cal Z}'\,,\label{HC}
\ea
where ${\cal Z}'$ is the generator of $U(1)_{Z'}$ in~(\ref{aas}). 
Under this definition the hypercharge assignments of the ordinary quarks and lepton fields are not altered. In contrast, 
 the hypercharge of the  states  
\ba
{\cal H}_{\cal L} &=&(1,3,1)=\hat
h_L^{+}\left(1,2;-\frac{1}{2},0\right)+\hat\nu_{{\cal H}_{\cal
L}}\left(1,1;0,0\right)\label{NSM1}
\\
{\cal H}_{\cal R} &=&(1,1,3)= {\hat
e_H^c}(1,1;1,0)+{\hat\nu_{{\cal H}_{\cal
R}}^{c+}}(1,1;0,1)+\hat\nu_{{\cal H}_{\cal R}}^{c-}(1,1;0,-1)\,,\label{NSM2}
\ea
have  non-zero components under $U(1)_{Z'}$ defined in~(\ref{aas}),
and  it turns out  that, with respect to~(\ref{HC}), the
states (\ref{NSM1},\ref{NSM2}) now carry  the SM electric charges.  Therefore,  provided
 the $U(1)_{Z'}$ gauge boson remains massless down to the electroweak scale, 
 the exotic  fractional states do not appear in this case.
The  mass of $Z'$ is affected   by higher dimensional anomalies and whether it  becomes massive or not,
depends on the details of the particular construction. 

Indeed,  let $D_a, D_b$ represent two stacks of the intersecting branes, 
where the topology of the 6-dimensional compact space is factorised into three tori ${\cal T}_i,\; i=1,2,3$.
Then, the multiplicities of the chiral fermions decending from the $D_a$-$D_b$ bifundamentals
are associated with the number of intersections 
\ba 
I_{ab}=\prod_{i=1}^3 (m_{ai}n_{bi}-m_{bi}n_{ai})~,\label{Iab}
\ea 
where $(n_{ai},m_{ai})$ are the winding numbers of the $D_a$ stack wrapping the two radii of the $i$-th torus. 
 Similar formulae can be written for fields arising from  other sectors. The restrictions on the $n_{ai},m_{ai}$ 
 winding numbers originating from the RR-type tadpole conditions  can be readily satisfied. The mixed 
anomalies $SU(3)_a^2\times U(1)_a$ are proportional to $I_{ab}$ and
impose additional restrictions on the $n_{ai},m_{ai}$ sets. 
For instance, after dimensional reduction the ten-dimensional fields
$C_2, C_6$ give the two-form fields $C_2= B_0$ and 
$B_2^i= \int_{{\cal T}_j\times {\cal T}_k}C_6$, and similar formulae
hold for their duals.

The coefficients involved in the anomaly cancellation conditions depend on  the winding numbers.
 The coefficient $c_a^0=m_{a1}m_{a2}m_{a3}$, in particular, couples directly to the linear combination~(\ref{HC}) through
  $B_2^0\wedge (F_c+ F_l+F_r)$ where $F_{c,l,r}$
are the corresponding field strengths associated with these three $U(1)$'s. 
In general, both anomaly cancellation and fermion multiplicities require $c_a^0\ne 0$, and, as a result the corresponding 
gauge boson $Z'$ becomes massive. In such a case, the new hypercharge definition cannot be implemented and so
 the states (\ref{NSM1},\ref{NSM2}) remain with exotic fractional charges.

\subsection{Gauge Couplings, Weak Mixing Angle and Monopole Mass in D brane Trinification}
 
The various stages of the symmetry breaking chain in the D-brane trinification
model are as follows. Initially, recall that for each brane stack $U(3)\simeq SU(3)\times U(1)$.
The $SU(3)_{L,R}$ symmetries are assumed to break at some intermediate scale between 
the $Z$ boson mass $M_Z\approx 92$ GeV and unification scale $M_{U}$.
The linear combination $U(1)_{Z'}$  may also break at any scale $M_{Z'}< M_{U}$.
However, if it is part of the hypercharge generator, this breaking should occur at low energies.

In the present trinification version the three gauge couplings $\alpha_{L,R,C}$ associated
with the three sets of  D-brane stacks are not necessarily equal. Hence,
in principle, there is enough freedom to reconcile the low energy values
of the gauge couplings with the experimental measurements.
Partial unification may lead to some constraints for the intermediate breaking scales.
In the most general scenario, we may assume that the gauge couplings of  $U(1)_{C,L,R}$
differ from those of the corresponding $SU(3)$  factors (perhaps due to threshold effects, etc).
Thus, we designate them with $\alpha_{L'},\alpha_{R'},\alpha_{C'}$.

 The generalized  hypercharge  embedding implies
 \be
 \frac{1}{\alpha_Y}= \frac{1}{3} \frac{1}{\alpha_L} +\frac{4}{3} \frac{1}{\alpha_R} +
 \kappa \frac{1}{6}\left( \frac{1}{\alpha_{L'}}+\frac{1}{\alpha_{R'}}+\frac{1}{\alpha_{C'}}\right)\,,
 \label{NHY}
 \ee
where  $\kappa =1$  for the general case, while for $\kappa =0$ we obtain the standard hypercharge assignment.
It is convenient to define the `harmonic' average
\be
    \frac{1}{\alpha_N} =  \frac{1}{3}\left( \frac{1}{\alpha_{L'}}+\frac{1}{\alpha_{R'}}+\frac{1}{\alpha_{C'}}\right)\,,
\ee
such that
\be
\sin^2\theta_{W} = \frac{3}{ 4\left(1+\frac{\alpha_L}{\alpha_{R}}\right)+\kappa\frac 32\frac{\alpha_L}{\alpha_N} },\;
\kappa= 0, 1\,.
  \ee
For $\kappa=0$ and $\alpha_L=\alpha_R$ we obtain the standard definition and the value $\sin^2\theta_W(M_U)=\frac 38$ at the GUT scale.
For  $\alpha_{L'}=\alpha_L=\alpha_{R'}=\alpha_R=\alpha_{C'}=\alpha_C$  and $\kappa=1$,  $\sin^2\theta_W(M_U)=\frac{6}{19}$.
For  $\alpha_{L'}\ne \alpha_L, \alpha_{R'}\ne \alpha_R$ etc., the standard  $\sin^2\theta_W(M_U)=\frac 38$ is
obtained if the condition  $\frac{1}{\alpha_R}+\frac{3}{8}\frac{1}{\alpha_N}=\frac{1}{\alpha_L}$  is fulfilled.
Notice however, that although in a general D-brane configuration such states are possible~\cite{Anastasopoulos:2006da},
in a minimal intersecting  D-brane scenario with just three brane stacks,  
the requirement for three fermion families imply~\cite{Leontaris:2008mm} a GUT mass for the gauge boson of the 
anomaly free $U(1)_{Z'}$  combination~(\ref{aas}). In such a case this cannot be used to modify the hypercharge generator
and as a result, the  representations $(1,3,1)$ etc. remain with fractional electric charges. 
For our purposes, in search for lighter monopoles and assuming trinification breaking not too far
from the EW scale, from eq (\ref{NHY}) setting $\kappa =0$, we find that $g_R\approx \sqrt{2}\,e$.

In the D-brane models a low unification scale is a plausible scenario since there 
is no compelling reason that the couplings unify at a high scale. As an illustrative example,
let us see how this works in the present case.  Let us designate the trinification scale
with $M_R$ and define the following combination at some scale $M_X\le M_R$:
\be 
\frac{1}{A_X}= \left(\frac{6}{\alpha_Y}-\frac{12}{\alpha_2}-\frac{1}{\alpha_3}\right)_{M_X}\,.
\ee
At $M_X=M_R$, it holds  $\alpha_2=\alpha_L, \alpha_3=\alpha_C$  while for the  hypercharge we use  formula (15) for $\kappa=1$ 
and $\alpha_i'=\alpha_i$~(a similar analysis can be easily performed for $\kappa=0$).
Also,  for mass scales $M_X$ in the energy scales between  $M_R$ and $M_U$, 
where $M_U$ is the GUT scale, the $SU(3)_C$ gauge coupling  
is eliminated in this  combination, so that 
\be 
\frac{1}{A_X}=9\left( \frac{1}{\alpha_L}-\frac{1}{\alpha_R}\right),\; \; {\rm for}\;\; M_R\le M_X\le M_U\,.
\ee
Then, the  trinification breaking scale is independent of the $a_C$ coupling, thus the latter 
can be fixed independently in order to give the known low energy value for $\alpha_3$.  
We can use now the Renormalization Group Equations (RGEs) to determine the trinification breaking
scale as a function of the known low energy values of the gauge couplings and beta functions. 
Matching the RGEs above and below the $M_R$ scale  we find that  
is given by
\be
M_R =e^{\frac{2\pi}{\beta-\beta'}\left(\frac{1}{A_Z}-\frac{1}{A_U}\right)}
\left(\frac{M_U}{M_Z}\right)^{\frac{\beta'}{\beta'-\beta}}\,M_Z\,,
\ee 
where $A_Z$ is given by $ A_X$ when evaluated at $ M_Z$  and $A_U$ = $A_X$ at 
$M_X = M_R$. Also, the coefficients  $\beta, \beta'$ are given by
\be
 \beta =6b_Y-12b_2-b_3\,,\; \;\beta' =9 (b_R-b_L)\,.
\ee
\begin{figure}
		\includegraphics[width=100mm]{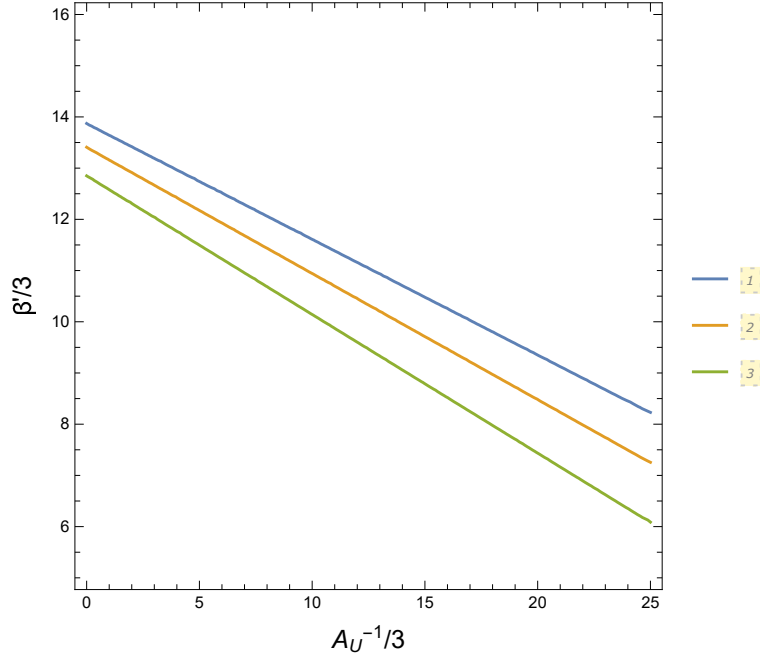}
	\caption{Contours for trinification breaking scale. The abscissa
		represents values for $({3A_U})^{-1}$
		which is proportional to the difference of the inverse gauge couplings $\alpha_{L,R}$
		at the	unification scale ${M_U}$~(see text). On the ordinate are the values for the difference	of the corresponding beta functions	$\beta'/3$.	 Curve (1) corresponds to 
		$M_R\sim 10^4$ GeV, curve (2)  to 
		$M_R\sim 10^5$ GeV, and curve (3)  to 
		$M_R\sim 10^6$ GeV.}
\end{figure}
For the particular case $b_L=b_R$, we get partial unification $\alpha_L=\alpha_R$
and,  since then $\beta'=0$,   the scale $M_R$ does not depend on $M_U$  and is fixed only in terms of the low energy parameters. We obtain
\be
M_R = e^{\frac{2\pi}{\beta}\frac{1}{A_Z}}M_Z\approx 7\times 10^{10}{\rm GeV}\,.
\ee 
In general, however, the string boundary conditions imply $b_L\ne b_R$ and therefore
various posibilities emerge. In Figure (2) we show contour plots for 
$M_R= 10^4, 10^5, 10^6$ GeV in the parameter space $\frac{1}{A_U}, b_R-b_L$.
For reasonable values $\frac{1}{\alpha_{L,R}}\sim {\cal O}(10)$ and $\beta' \propto b_R-b_L$,
 the trinification scale can be as low as $10^4$ to $10^6$ GeV.

The reader might wonder whether a low trinification breaking scale could have catastrophic consequences for baryon number violating processes. 
Firstly, we recall that trinification symmetry does not contain gauge boson mediated dimension six proton decay operators. Secondly, as we
have already pointed out, all baryon fields ${\cal Q} = (3, \bar 3, 1)$ carry the same charge under the abelian
symmetry $U(1)_C$ and, therefore, the latter could play the r\^ole of baryon number. Finally,
introducing a suitable `matter' parity in order to distinguish the Higgs and lepton multiplets,
${\cal H}={\cal L} = (1, 3, \bar 3)$, the only allowed Yukawa coupling involving the quark fields is ${\cal Q} {\cal Q}^c{\cal H}$. Thus,
proton decay can be adequately suppressed in this class of trinification models.

Before closing this section, we point out that a similar intersecting D-brane configuration can be arranged for
the 422 model where states with fractional charges $\pm \frac e6,\, \pm \frac e2$ are
generated by open strings with appropriate boundary conditions.  The states with electric charges $\pm \frac e6$ 
also carry color and are therefore confined. 

\section{Monopoles, Inflation and Primordial Gravity Waves:}

Magnetic monopoles can be problematic in the standard big bang cosmology. If they are produced
at a high, $M_{U}\sim 10^{16}$GeV unification scale where a $U(1)$ emerges from a non-abelian gauge group, then they
overclose the Universe in the standard hot big-bang cosmology. This problem is solved by inflation which dilutes the monopoles, in some cases to levels that agree with observation.
Then, it is perfectly reasonable to ask the question: how do primordial monopoles survive cosmic inflation?

This has been addressed in a number of ways by various authors and we very briefly summarize a few of them.
Firstly, suppose that the spontaneous breaking of non-supersymmetric $SO(10)$ to the SM proceeds via the 422 symmetry, with inflation driven by an SO(10) singlet
field~\cite{Lazarides:1984pq} using the Coleman-Weinberg potential. For this case a scalar spectral index $n_s\sim 0.96-0.97$ is realized for a Hubble constant $H_{inf}$ during inflation of order $10^{13} -10^{14}$ GeV\cite{Senoguz:2015lba}. This leads to the conclusion that monopoles associated with the breaking of 422 at an energy scale close to $H_{inf}$ can survive the inflationary epoch and be present in our galaxy at an observable level. This $SO(10)$ inflationary scenario also predicts that the tensor to scalar ratio $r$, a canonical measure of gravity waves, is not much
 smaller than 0.02 \cite{Rehman:2008qs}, which will be tested in the near future.

A somewhat different inflationary scenario based on a quartic potential with non-minimal coupling of the inflaton field to gravity predicts an $r$ value up to an order of magnitude or so smaller~\cite{Okada:2010jf} than the previous example. The monopole mass in this case is around $10^{13}$--$10^{14}$ GeV.

Monopoles arising in models such as supersymmetric   trinification have been shown~\cite{Lazarides:1986rt} to survive primordial inflation by exploiting an epoch of thermal inflation~\cite{Lazarides:1987zf}-\cite{Lyth:1995ka} which dilutes their number density to levels below the Parker bound. Depending on the model details the monopole masses can vary from the intermediate to GUT scale.

If the theory has a product group that avoids proton decay without being
 broken at a high scale and if the monopoles are not produced
until near the electroweak scale, then they could be eliminated by late time
inflation, although this may not be easy to arrange. Another possibility
 \cite{Langacker:1980kd,Farris:1991rg} is to eliminate them or substantially reduce their numbers
 by temporarily breaking the appropriate $U(1)$. Then the monopoles
find themselves on the end of cosmic strings. The high tension in the strings
causes efficient monopole-antimonopole annihilation thereby solving the cosmic
monopole problem. Either of these mechanisms allows one to bring the monopole mass
density down to a value that does not conflict with present astrophysical
observations.

\section{Discussion}

  Here we explore the possibility of detection of low and intermediate mass magnetic monopoles, especially those that are multiply charged.
For the detection of low mass magnetic monopoles we focus on the LHC, and for the detection of intermediate mass magnetic monopoles we focus on
cosmic ray experiments.

  \subsection{Monopoles at the LHC}

There have been recent suggestions of light monopoles in the standard model~\cite{Cho:1996qd,Cho:2012bq,Vento:2013jua,Cho:2013vla} and this possibility can also be explored in various branches of the 433 model. Singly charged monopoles (i.e., charge $n=1$) interact strongly with matter \cite{Kephart:1995bi,Wick:2000yc} through their fine structure constant
$$\alpha_M=\frac{1}{2\alpha}\sim68\,,$$
and cross sections are enhanced by a factor of $n^2$ for multiply charged monopoles. Hence the reach of the LHC is long if it  produces $M\overline{M}$ pairs. But as we discuss below, production at the LHC requires that the $M\overline{M}$ pairs are fundamental, i.e., of Dirac type, 
since 't Hooft-Polyakov~\cite{'tHooft:1974qc,Polyakov:1974ek} monopoles, being composite, are much harder to produce and their production cross section has been estimated to be suppressed by greater than 30 orders of magnitude relative to production of fundamental point-like monopoles~\cite{Drukier:1981fq}. Hence, composite monopoles are extremely unlikely to be accessible at the LHC. (For other possibilities see also~\cite{Zeldovich:1978wj},\cite{Stojkovic:2007dw})

Above threshold fundamental $M\overline{M}$ pairs will be copiously produced and easily detected by their densely ionizing tracks in detectors.
The MoEDAL experiments \cite{Pinfold:2010zza, MoEDAL:2016jlb,King:2016pys,Pinfold:2016jji,Acharya:2016ukt,Mavromatos:2016ykh} searches for monopoles both by tracking in layered material and by monopole capture in aluminum bares that are run through superconducting detectors. Both these types of searches are carried out offline.

Monopoles will be accelerated (or decelerated) in
detector magnets, and will travel on parabolic trajectories in constant magnetic fields. Hence their track will not look at all like electrically charged particles traveling on helical orbits in magnetic fields. Track reconstruction fitting routines can easily be made to distinguish the difference. Combining ionization with tracking could make a monopole track even more unmistakable.

Below threshold virtual pairs of monopoles can contribute to loop diagrams for scattering processes and alter cross sections from their predicted SM values. For example, Drell-Yan like production cross sections  $q \bar q \rightarrow X\overline{X}$ could be enhanced.

Even though production cross sections of composite 't Hooft-Polyakov monopoles  are too small for them to be produced
at accelerators, this is not the case for point-like Dirac monopoles. While a full quantum theory of magnetic monopoles is lacking,
limits on Dirac monopole production have been obtained via a Drell-Yan model \cite{Kalbfleisch:2003yt},
and applied to monopoles of 1, 2, 3 and 6 times the Dirac charge for 175 pb$^{-1}$ exposure of $p\bar{p}$ luminosity of material in the collision regions of both D0 and CDF. The resulting monopole mass limits are 256, 355, 410 and 375 GeV/$c^2$ respectively, while the production cross section limits are 0.6, 0.2, 0.07 and 0.2 pb respectively.

The fractional electric charges are also very interesting in these models and potentially detectable. The electric charges are often in multiples
of $\frac{1}{2}e$ or $\frac{1}{3}e$, but other fractions of $e$ are possible in certain embeddings of the SM in the 433 model. In one case particle charges come in fractions as small as $\frac{1}{12}e$~\cite{Kephart:2006zd}.

In the past, many of the best magnetic monopole limits~(a comprehensive list of references can be found in the  `Magnetic Monopole Bibliography,'' of Giacomelli et al., \cite{Giacomelli:2000de,Balestra:2011ks},)  have been based on cosmic ray experiments   \cite{Hogan:2008sx,Christy:2011lza,AdrianMartinez:2011xr,Popa:2011zza,Balestra:2008ps}, but now there is also a dedicated experiment at the LHC for this purpose.  The MoEDAL experiment \cite{Pinfold:2010zza, MoEDAL:2016jlb,King:2016pys,Pinfold:2016jji,Acharya:2016ukt,Mavromatos:2016ykh} mentioned above has been  specifically designed  to search  for  magnetic monopoles and other highly ionizing particles. The ATLAS experiment has also reported on their magnetic monopole search \cite{ATLAS:2012nla}. We hope the results presented here can provide additional motivation to these and other experiments.

\subsection{Monopoles in Cosmic Rays}   The number density  of monopoles emerging from
an early universe phase transition is determined by the Kibble mechanism \cite{Kibble:1976sj,Kibble:1980mv}. From the number density we
determine the flux of free monopoles with $M < 10^{15}$~GeV  accelerated to relativistic energies by the cosmic magnetic fields.
The general expression for the relativistic monopole flux may be written \cite{Kephart:1995bi,Wick:2000yc}
\be
F_M = c\: n_M/4\pi
 \sim 2\times 10^{-4}\, \left(\frac{M}{{10^{15}{\rm GeV}}}\right)^3
\left(\frac{l_H}{\xi_c}\right)^3\,
{\rm cm}^{-2} \;{\rm sec}^{-1}\;{\rm sr}^{-1}\,.
\label{flux}
\ee

The IceCube experiment has recently put a limit on the flux of light mildly relativistic ($\beta < 0.8$) magnetic monopoles \cite{Abbasi:2012eda,Aartsen:2015exf}
 $$\Phi_{90\% C.L.} \sim 10^{-18}cm^{-2}sr^{-1}s^{-1}\,.$$
This in turn limits the cosmic density of magnetic monopoles, but it does not eliminate the possibility that cosmic monopoles were all  either inflated away or annihilated at the electroweak scale but can now still be produced in accelerator or cosmic ray collisions if they are point-like particles.

Magnetic monopoles in cosmic rays could have been produced in the early universe and therefore could be of either composite 't Hooft-Polyakov type or Dirac point-like type.    The Pierre Auger experiment has recently reported on a search for ultra relativistic magnetic monopoles \cite{Aab:2016poe} and placed limits on their flux of $1 \times 10^{-19}$ (cm$^2$ sr s)$^{-1}$ and $2.5 \times 10^{-21}$ (cm$^2$ sr s)$^{-1}$ for Lorentz factors of $\gamma = 10^9$ and $\gamma = 10^{12}$ respectively, and as mentioned above, IceCube has also placed limits on the flux of relativistic and mildly relativistic magnetic monopoles \cite{Aartsen:2015exf}. For velocities above 0.51 $c$ they see no flux above $1.55 \times 10^{-18}$ (cm$^2$ sr s)$^{-1}$. The best upper limit on the flux of nonrelativistic magnetic monopoles comes from MACRO \cite{Ambrosio:2002qq} who find $1.5 \times 10^{-16}$ (cm$^2$ sr s)$^{-1}$ for $4 \times 10^{-5}<\beta = v/c<0.5 $, where all fluxes above are quoted at the 90\% C.L. Numerous other experiments have also placed limits on the flux of magnetic monopoles in cosmic rays Baikal \cite{Aynutdinov:2005sg}, SLIM \cite{Balestra:2008ps}, RICE \cite{Hogan:2008sx} and ANITA-II \cite{Detrixhe:2010xi}, with the best limit on ultra relativistic magnetic monopoles coming from ANITA-II and the best limit on  relativistic magnetic monopoles $\beta =0.9$ coming from IceCube \cite{Abbasi:2012eda}, as discussed and summarized in \cite{Aab:2016poe}.

\begin{acknowledgments}
Q.S. is supported in part by the DOE Grant DE-SC0013880.
T.W.K and G.K.L. would like to thank the Physics and Astronomy Department and Bartol Research
Institute of the University of Delaware for kind hospitality. Also, G.K.L. would like to thank
LPTHE of  UPMC in Paris for kind hospitality during the final stages of this work.

\end{acknowledgments}


\begin{thebibliography}{100}
\bibitem{Dirac}
  P.~A.~M.~Dirac,
  Proc.\ Roy.\ Soc.\ Lond.\ A {\bf 133}, 60 (1931).




\bibitem{PS} J. C. Pati and A. Salam, Phys. Rev. D 10, 275 (1974).

\bibitem{Tri} S. L. Glashow, FIFTH WORKSHOP ON GRAND UNIFICATION: proceedings. Edited
by Kyungsik Kang, Herbert Fried, Paul Frampton, (World Scientific, 1984) 538p.

\bibitem{Lazarides:1980cc}
  G.~Lazarides, M.~Magg and Q.~Shafi,
  Phys.\ Lett.\ B {\bf 97}, 87 (1980).

\bibitem{Shafi:2007xp}
  Q.~Shafi and C.~A.~Lee,
  Phys.\ Lett.\ B {\bf 661}, 33 (2008)
  [arXiv:0709.4637 [hep-ph]].

\bibitem{Lazarides:1982jq}
  G.~Lazarides, Q.~Shafi and W.~P.~Trower,
  Phys.\ Rev.\ Lett.\  {\bf 49}, 1756 (1982).

\bibitem{GQW}
  H.~Georgi, H.~R.~Quinn and S.~Weinberg,
  Phys.\ Rev.\ Lett.\  {\bf 33}, 451 (1974).

\bibitem{Georgi:1974sy}
  H.~Georgi and S.~L.~Glashow,
  Phys.\ Rev.\ Lett.\  {\bf 32}, 438 (1974).

\bibitem{Lawrence:1998ja}
  A.~E.~Lawrence, N.~Nekrasov and C.~Vafa,
  Nucl.\ Phys.\ B {\bf 533}, 199 (1998)
  [hep-th/9803015].

\bibitem{Ho:2011qi}
  C.~M.~Ho, P.~Q.~Hung and T.~W.~Kephart,
  JHEP {\bf 1206}, 045 (2012)
  [arXiv:1102.3997 [hep-ph]].

  
  \bibitem{extraD}
  K.~R.~Dienes, E.~Dudas and T.~Gherghetta,
  Phys.\ Lett.\ B {\bf 436} (1998) 55
  [hep-ph/9803466].
  \\
  L.~J.~Hall and Y.~Nomura,
  Phys.\ Rev.\ D {\bf 64} (2001) 055003
  [hep-ph/0103125].
  \\
  G.~Shiu and S.~H.~H.~Tye,
  Phys.\ Rev.\ D {\bf 58} (1998) 106007
  [hep-th/9805157].
  

\bibitem{Dvali:1994wj}
  G.~R.~Dvali and Q.~Shafi,
  Phys.\ Lett.\ B {\bf 326}, 258 (1994)
  [hep-ph/9401337].

\bibitem{Berezhiani:2001ub}
  Z.~Berezhiani, I.~Gogoladze and A.~Kobakhidze,
  Phys.\ Lett.\ B {\bf 522}, 107 (2001)
  [hep-ph/0104288].

\bibitem{Barr:2014nza}
  S.~M.~Barr and X.~Calmet,
  JHEP {\bf 1407}, 159 (2014)
  [arXiv:1404.4594 [hep-ph]].


\bibitem{Kephart:2001ix}
 T.~W.~Kephart and Q.~Shafi,
  ``Family Unification, Exotic States and Magnetic Monopoles,''
  Phys.\ Lett.\ B {\bf 520}, 313 (2001)
  [hep-ph/0105237].

\bibitem{Kephart:2006zd}
  T.~W.~Kephart, C.~-A.~Lee and Q.~Shafi,
  ``Family Unification, Exotic States and Light Magnetic Monopoles,''
  JHEP {\bf 0701}, 088 (2007)
  [hep-ph/0602055].
  
  
  \bibitem{Cho:1996qd}
  Y.~M.~Cho and D.~Maison,
  ``Monopoles in Weinberg-Salam model,''
  Phys.\ Lett.\ B {\bf 391}, 360 (1997)
  [hep-th/9601028].
  
  \bibitem{Cho:2012bq}
  Y.~M.~Cho, K.~Kimm and J.~H.~Yoon,
  ``Mass of the Electroweak Monopole,''
  Snowmass on the Mississippi Workshop (CSS2013),
  arXiv:1212.3885 [hep-ph].
  
  \bibitem{Vento:2013jua}
  V.~Vento and V.~S.~Mantovani,
  ``On the magnetic monopole mass,''
  arXiv:1306.4213 [hep-ph].
  
  \bibitem{Cho:2013vla}
  Y.~M.~Cho and J.~L.~Pinfold,
  ``Electroweak Monopole Production at the LHC - a Snowmass White Paper,''
  arXiv:1307.8390 [hep-ph].
  

\bibitem{Anastasopoulos:2006da}
  P.~Anastasopoulos, T.~P.~T.~Dijkstra, E.~Kiritsis and A.~N.~Schellekens,
  Nucl.\ Phys.\ B {\bf 759} (2006) 83
  [hep-th/0605226].
    G.~K.~Leontaris and J.~Rizos,
    Phys.\ Lett.\ B {\bf 632}, 710 (2006).
    
    
    \bibitem{Leontaris:2008mm}
    G.~K.~Leontaris,
    Int.\ J.\ Mod.\ Phys.\ A {\bf 23} (2008) 2055
    doi:10.1142/S0217751X0804055X
    [arXiv:0802.4301 [hep-ph]].

\bibitem{Ibanez:1998qp}
L.~E.~Ibanez, R.~Rabadan and A.~M.~Uranga,
Nucl.\ Phys.\ B {\bf 542} (1999) 112
doi:10.1016/S0550-3213(98)00791-3
[hep-th/9808139].
G.~Aldazabal, S.~Franco, L.~E.~Ibanez, R.~Rabadan and A.~M.~Uranga,
J.\ Math.\ Phys.\  {\bf 42} (2001) 3103
doi:10.1063/1.1376157
[hep-th/0011073].


\bibitem{Lazarides:1984pq}
G.~Lazarides and Q.~Shafi,
Phys.\ Lett.\  {\bf 148B} (1984) 35.

\bibitem{Senoguz:2015lba}
V.~N. Senoguz and Q.~Shafi,
Phys.\ Lett.\ B {\bf 752} (2016) 169
[arXiv:1510.04442 [hep-ph]].


\bibitem{Rehman:2008qs}
M.~U.~Rehman, Q.~Shafi and J.~R.~Wickman,
Phys.\ Rev.\ D {\bf 78} (2008) 123516
[arXiv:0810.3625 [hep-ph]].


\bibitem{Okada:2010jf}
N.~Okada, M.~U.~Rehman and Q.~Shafi,
Phys.\ Rev.\ D {\bf 82} (2010) 043502
[arXiv:1005.5161 [hep-ph]].


\bibitem{Lazarides:1986rt}
G.~Lazarides, C.~Panagiotakopoulos and Q.~Shafi,
Phys.\ Rev.\ Lett.\  {\bf 58} (1987) 1707.


\bibitem{Lazarides:1987zf}
G.~Lazarides, C.~Panagiotakopoulos and Q.~Shafi,
Phys.\ Lett.\ B {\bf 192} (1987) 323.


\bibitem{Lazarides:1985ja}
G.~Lazarides, C.~Panagiotakopoulos and Q.~Shafi,
Phys.\ Rev.\ Lett.\  {\bf 56} (1986) 557.


\bibitem{Yamamoto:1985rd}
K.~Yamamoto,
Phys.\ Lett.\  {\bf 168B} (1986) 341.


\bibitem{Binetruy:1986ss}
P.~Bin\'etruy and M.~K.~Gaillard,
Phys.\ Rev.\ D {\bf 34} (1986) 3069.


\bibitem{Lazarides:1987yq}
G.~Lazarides, C.~Panagiotakopoulos and Q.~Shafi,
Nucl.\ Phys.\ B {\bf 307} (1988) 937.

\bibitem{Lazarides:1992gg}
G.~Lazarides and Q.~Shafi,
Nucl.\ Phys.\ B {\bf 392} (1993) 61.

\bibitem{Lyth:1995ka}
D.~H.~Lyth and E.~D.~Stewart,
Phys.\ Rev.\ D {\bf 53} (1996) 1784
[hep-ph/9510204].


\bibitem{Langacker:1980kd}
  P.~Langacker and S.~-Y.~Pi,
  Phys.\ Rev.\ Lett.\  {\bf 45}, 1 (1980).

\bibitem{Farris:1991rg}
  T.~H.~Farris, T.~W.~Kephart, T.~J.~Weiler and T.~C.~Yuan,
  Phys.\ Rev.\ Lett.\  {\bf 68}, 564 (1992).


\bibitem{Kephart:1995bi}
  T.~W.~Kephart and T.~J.~Weiler,
  Astropart.\ Phys.\  {\bf 4}, 271 (1996)
  [astro-ph/9505134].

\bibitem{Wick:2000yc}
  S.~D.~Wick, T.~W.~Kephart, T.~J.~Weiler and P.~L.~Biermann,
  Astropart.\ Phys.\  {\bf 18}, 663 (2003)
  [astro-ph/0001233].
  
  
\bibitem{'tHooft:1974qc}
  G.~'t Hooft,
  Nucl.\ Phys.\ B {\bf 79}, 276 (1974).

\bibitem{Polyakov:1974ek}
  A.~M.~Polyakov,
  JETP Lett.\  {\bf 20}, 194 (1974)
  [Pisma Zh.\ Eksp.\ Teor.\ Fiz.\  {\bf 20}, 430 (1974)].

\bibitem{Drukier:1981fq} 
  A.~K.~Drukier and S.~Nussinov,
  Phys.\ Rev.\ Lett.\  {\bf 49}, 102 (1982).
  
  \bibitem{Zeldovich:1978wj}
  Y.~B.~Zeldovich and M.~Y.~Khlopov,
  Phys.\ Lett.\  {\bf 79B} (1978) 239.
  doi:10.1016/0370-2693(78)90232-0
  
 \bibitem{Stojkovic:2007dw}
 D.~Stojkovic, G.~D.~Starkman and R.~Matsuo,
 Phys.\ Rev.\ D {\bf 77} (2008) 063006
 doi:10.1103/PhysRevD.77.063006
 [hep-ph/0703246].
  
  
\bibitem{Pinfold:2010zza}
  J.~Pinfold [MOEDAL Collaboration],
  ``MoEDAL becomes the LHC's magnificent seventh,''
  CERN Cour.\  {\bf 50N4}, 19 (2010).

\bibitem{MoEDAL:2016jlb}
  B.~Acharya {\it et al.} [MoEDAL Collaboration],
  JHEP {\bf 1608}, 067 (2016)
  [arXiv:1604.06645 [hep-ex]].
  
  

\bibitem{King:2016pys}
  M.~King [MoEDAL Collaboration],
  Nucl.\ Part.\ Phys.\ Proc.\  {\bf 273-275}, 2560 (2016).

\bibitem{Pinfold:2016jji}
  J.~L.~Pinfold [MoEDAL Collaboration],
  EPJ Web Conf.\  {\bf 126}, 02024 (2016).
  


\bibitem{Acharya:2016ukt}
B.~Acharya {\it et al.} [MoEDAL Collaboration],
Phys.\ Rev.\ Lett.\  {\bf 118} (2017) no.6,  061801
[arXiv:1611.06817 [hep-ex]].



\bibitem{Mavromatos:2016ykh}
  N.~E.~Mavromatos {\it et al.} [MoEDAL Collaboration],
  arXiv:1612.07012 [hep-ph].

\bibitem{Giacomelli:2000de}
  G.~Giacomelli, M.~Giorgini, T.~Lari, M.~Ouchrif, L.~Patrizii, V.~Popa, P.~Spada and V.~Togo,
  hep-ex/0005041.

\bibitem{Balestra:2011ks}
S.~Balestra, G.~Giacomelli, M.~Giorgini, L.~Patrizii, V.~Popa, Z.~Sahnoun and V.~Togo,
arXiv:1105.5587 [hep-ex].

\bibitem{Christy:2011lza}
  B.~J.~Christy,
  ``A Search for Relativistic Magnetic Monopoles with the IceCube 22-String Detector,''
  Ph. D. dissertation, Maryland U., College Park.

\bibitem{AdrianMartinez:2011xr}
  S.~Adrian-Martinez {\it et al.}  [ANTARES Collaboration],
  ``Search for Relativistic Magnetic Monopoles with the ANTARES Neutrino Telescope,''
  Astropart.\ Phys.\  {\bf 35}, 634 (2012)
  [arXiv:1110.2656 [astro-ph.HE]].

\bibitem{Popa:2011zza}
  V.~Popa [KM3NeT Collaboration],
  ``KM3NeT: Present status and potentiality for the search for exotic particles,''
  Nucl.\ Instr.\ Meth.\ A {\bf 630}, 125 (2011).
  
  \bibitem{Balestra:2008ps}
  S.~Balestra {\it et al.},
  Eur.\ Phys.\ J.\ C {\bf 55}, 57 (2008)
  [arXiv:0801.4913 [hep-ex]].
  
  \bibitem{Hogan:2008sx}
  D.~P.~Hogan, D.~Z.~Besson, J.~P.~Ralston, I.~Kravchenko and D.~Seckel,
  Phys.\ Rev.\ D {\bf 78}, 075031 (2008)
  [arXiv:0806.2129 [astro-ph]].
 
\bibitem{ATLAS:2012nla}
  [ATLAS Collaboration],
  ``Search for magnetic monopoles in $\sqrt{s}=7$~TeV $pp$ collisions with the ATLAS detector,''
  ATLAS-CONF-2012-062.

\bibitem{Kibble:1976sj}
  T.~W.~B.~Kibble,
  J.\ Phys.\ A {\bf 9}, 1387 (1976).

\bibitem{Kibble:1980mv}
  T.~W.~B.~Kibble,
  Phys.\ Rept.\  {\bf 67}, 183 (1980).
  
  \bibitem{Aartsen:2015exf}
  M.~G.~Aartsen {\it et al.} [IceCube Collaboration],
  Eur.\ Phys.\ J.\ C {\bf 76}, no. 3, 133 (2016)
  [arXiv:1511.01350 [astro-ph.HE]].

\bibitem{Abbasi:2012eda}
R.~Abbasi {\it et al.} [IceCube Collaboration],
Phys.\ Rev.\ D {\bf 87}, no. 2, 022001 (2013)
[arXiv:1208.4861 [astro-ph.HE]].

\bibitem{Kalbfleisch:2003yt}
  G.~R.~Kalbfleisch, W.~Luo, K.~A.~Milton, E.~H.~Smith and M.~G.~Strauss,
  Phys.\ Rev.\ D {\bf 69}, 052002 (2004)
  [hep-ex/0306045].


\bibitem{Aab:2016poe}
  A.~Aab {\it et al.} [Pierre Auger Collaboration],
  Phys.\ Rev.\ D {\bf 94}, no. 8, 082002 (2016)
  [arXiv:1609.04451 [astro-ph.HE]].

\bibitem{Ambrosio:2002qq}
  M.~Ambrosio {\it et al.} [MACRO Collaboration],
  Eur.\ Phys.\ J.\ C {\bf 25}, 511 (2002)
  [hep-ex/0207020].

\bibitem{Aynutdinov:2005sg}
  V.~Aynutdinov {\it et al.} [Baikal Collaboration],
  astro-ph/0507713.


\bibitem{Detrixhe:2010xi}
  M.~Detrixhe {\it et al.} [ANITA-II Collaboration],
  Phys.\ Rev.\ D {\bf 83}, 023513 (2011)
  [arXiv:1008.1282 [astro-ph.HE]].


\end{thebibliography}
\end{document}